# Parallel frequency function-deep neural network for efficient complex broadband signal approximation

Zhi Zeng[+], Pengpeng Shi[+*], Fulei Ma, Peihan Qi

[+] These authors contributed equally to this paper.

[*] **Corresponding author**

Pengpeng Shi is with the School of Civil Engineering & Institute of Mechanics and Technology in Xi'an University of Architecture and Technology, China

E-mail: shipengpeng@xjtu.edu.cn

**Abstract**

A neural network is essentially a high-dimensional complex mapping model by adjusting network weights for feature fitting. However, the spectral bias in network training leads to unbearable training epochs for fitting the high-frequency components in broadband signals. To improve the fitting efficiency of high-frequency components, the PhaseDNN was proposed recently by combining complex frequency band extraction and frequency shift techniques [Cai et al. SIAM J. SCI. COMPUT. 42, A3285 (2020)]. Our paper is devoted to an alternative candidate for fitting complex signals with high-frequency components. Here, a parallel frequency function-deep neural network (PFF-DNN) is proposed to suppress computational overhead while ensuring fitting accuracy by utilizing fast Fourier analysis of broadband signals and the spectral bias nature of neural networks. The effectiveness and efficiency of the proposed PFF-DNN method are verified based on detailed numerical experiments for six typical broadband signals.

# 1. Introduction

Artificial neural network, also known as the neural network, refers to a mathematical model that imitates animal neural network behavior [1], which is essentially a high-dimensional complex mapping model by adjusting network weights for feature fitting. A neural network's basic building block is a neuron model, a parameterized model nested by a scalar linear function, and a monotonic nonlinear function (activation function), where the coefficients in the linear function are the connection weights between neurons. Neurons connecting according to a specific topology forms a neural network. One of the primary networks is the single-layer network composed of multiple neurons in parallel. Multiple single-layer networks can be stacked to obtain a multi-layer network and further expanded into a deep neural network that contains multiple multi-layer networks. Some advanced neural network models are designed to meet the needs of engineering practice [2-5]. For example, Convolutional neural networks (CNN) and their evolution models have achieved unprecedented success in computer vision due to their powerful feature extraction capabilities, and are widely used in security monitoring, autonomous driving, human-computer interaction, augmented reality, and many other fields. And recurrent neural networks, especially the long-short-term memory model (LSTM), have become mainstream tools in the fields of automatic translation, text generation, and video generation in just a dozen years.

The related research of the universal approximation [6-8] indicates that assuming sufficient neurons and suitable weights, the neural network could approximate any continuous signals on a compact subset of $\mathbb{R}^n$ with arbitrary precision. However, it is not easy to obtain these appropriate weights via training for a complex network with too many neurons. The convergence speed of the neural network is related to the frequency spectrum of the fitted signal [9]. As shown in Fig. 1, the neural network first learns the low-frequency components during the training process. The relationship between the convergence speed and the frequency of the fitted signal has been quantitatively analyzed [10]. When a network is applied to fit a signal, the required training epochs increase exponentially as the component's central frequency increases. The spectral bias of the convergence speed in network training leads to unbearable training epochs for fitting the high-frequency components in broadband signals. To solve this problem, Cai et al. proposed the PhaseDNN method for fitting complex signals with high-frequency components via the combination of parallel frequency band extraction and frequency shifting techniques [11]. Numerous numerical experimental results indicate that PhaseDNN successfully avoids the computational cost disaster when fitting the signals with high-frequency components. However, the excessive training overhead still exists for many typical broadband signals because the extracted frequency bandwidth needs to be minimized to ensure fitting accuracy.

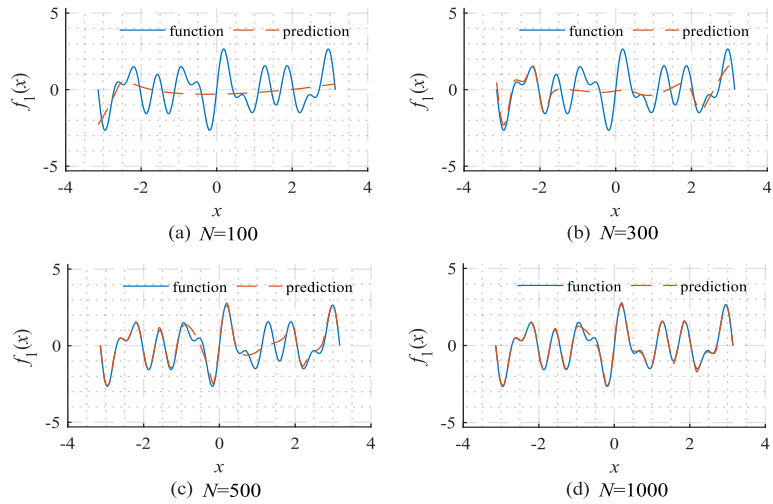

**Fig. 1.** Network fitting results given different number of updating reflects that neural networks often learn low-frequency components first in the training process: (a) $N=100$, (b) $N=300$, (c) $N=500$, (d) $N=1000$. The objective function for fitting is $f_1= \sin5x+\sin7x+\sin11x$. The network uses a 1-40-40-40-40-1 type fully connected structure. The Adam optimizer is used for optimization under the default learning rate.

Our paper is devoted to an alternative candidate for fitting complex signals with high-frequency components that can ensure fitting accuracy without causing a large number of the extracted frequency selected signals. Here, a parallel frequency function-deep neural network (PFF-DNN) is proposed by utilizing fast Fourier analysis of broadband signals and the spectral bias nature of neural networks. The effectiveness and efficiency of the proposed method are verified based on detailed experiments for six typical broadband signals. The discussion shows how the neural network adaptively constructs low-frequency smooth curves to fit discrete signals. This adaptive low-frequency approximation makes it possible to fit discrete frequency domain signals accurately.

## 2. The Methods

### 2.1. Recent PhaseDNN

The research mentioned above indicates that it is difficult for neural networks to directly and accurately fit signals with high-frequency components. But if we can transform the high-frequency components in the signal into smooth low-frequency signals convenient for neural network fitting by some means (such as frequency shift technology), we can approximate the original signal via component-by-component fitting. Following this idea, in 2020, Cai et al. proposed the PhaseDNN method to implement a fitting scheme for signals with high-frequency components [11]. As shown in Fig. 2, the PhaseDNN method involves four steps: 1) All high-frequency components in the original objective signal are extracted; 2) Each high-frequency component is converted into a low-frequency signal by using the frequency shift technology; 3) Different neural networks with the identical structure are used to fit these low-frequency signals in parallel; 4) Inverse frequency shift operation are performed for all network predictions to obtain approximated high-frequency components, which are further summed up to recover the original signal.

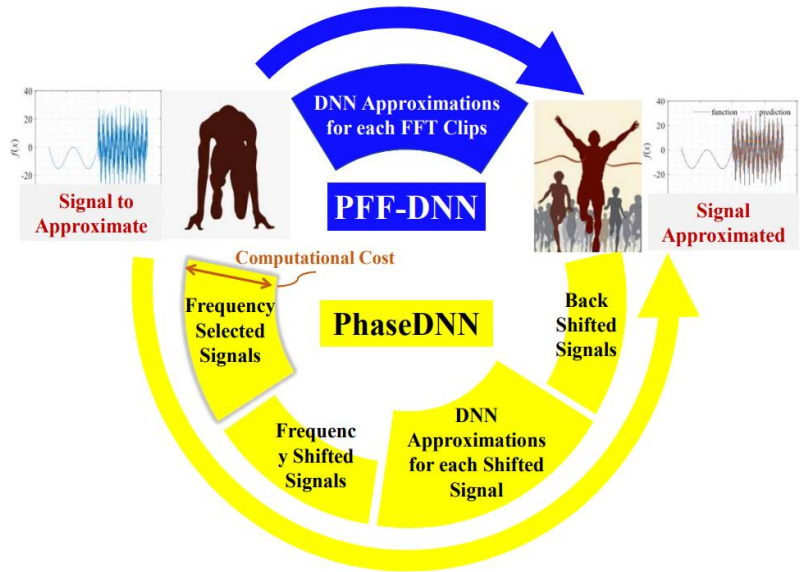

**Fig. 2.** Flowcharts of the Phase-DNN algorithms and the proposed PFF-DNN method.

Fig. 3 shows a comparison between the fitted results from PhaseDNN and that obtained via vanilla fitting. As shown in Fig. 3(a), vanilla fitting cannot recover high-frequency components well. If the signal oscillates even faster, the vanilla fitting becomes entirely ineffective. On the contrary, as shown in Fig. 3(c and d), the recent PhaseDNN method has shown obvious advantages in completely characterizing the information of all frequency bands of the signal.

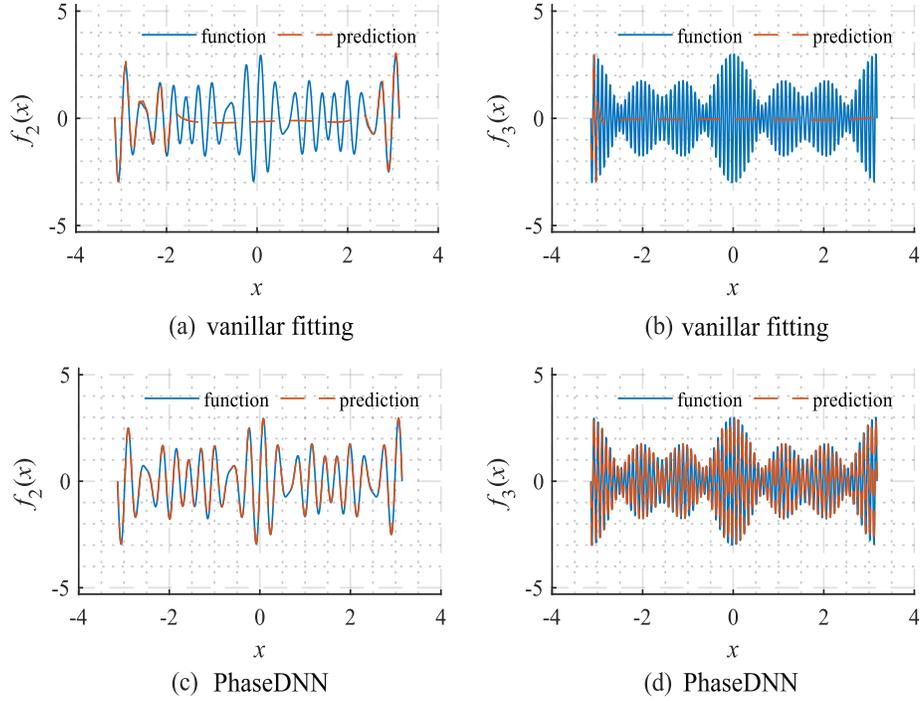

**Fig. 3.** Comparison between the fitted results from PhaseDNN and that obtained by vanilla fitting when $N=1000$. The objective functions for fitting are $f_2 = \sin17x+\sin19x+\sin23x$ and $f_3 = \sin67x+\sin71x+\sin73x$. (a) Fitting $f_2$ via vanilla fitting; (b) Fitting $f_3$ via vanilla fitting; (c) Fitting $f_2$ via PhaseDNN; (d) Fitting $f_3$ via PhaseDNN. The network uses a 1-40-40-40-40-1 type fully connected structure. The Adam optimizer is used for optimization under the default learning rate.

It should be noted that the extracted frequency bandwidth needs to be minimized to ensure fitting accuracy. On the one hand, the smaller the extracted frequency band's width $\Delta\omega$, the higher the fitting accuracy. On the other hand, a smaller $\Delta\omega$ indicates more frequency bands are extracted, which directly leads to an increase in computational overhead. (When using PhaseDNN for broadband signal fitting, it is often necessary to extract all frequency bands and fit them separately.) Therefore, there is a balance between accuracy and computational overhead. For example, the task is to perform neural network fitting on a signal with a bandwidth of 300 Hz. When $\Delta\omega=10$, considering the existence of both real and imaginary parts in the spectrum and the conjugate symmetry, 30×2 groups of neural networks need to be trained, which is an acceptable computational overhead. However, for the task of neural network fitting to a signal with a bandwidth of 3000 Hz, 3000×2 groups of neural networks need to be trained, which significantly increases the computational overhead.

## 2.2. The Proposed PFF-DNN Method

It is preferable if we have such a method that can ensure the fitting accuracy without causing a large number of the extracted frequency selected signals. Using fast Fourier transformation (FFT) of broadband signals [12], one can get the digital spectrum of the signal much efficiently. It is conceivable that if we perform a piecewise fitting on a signal's digital spectrum, no computational overhead is required for frequency selection nor frequency shift. This avoids several redundant operations in the PhaseDNN method. In addition, when the frequency spectrum of the signal is not overcomplicated, fitting in the frequency domain will significantly improve efficiency. The method proposed here is shortened as the PFF-DNN, and the details of it are as follows:

The objective signal is denoted as a real-valued $f(x)$ in the domain of $[x_0, x_{n-1}]$. From a digital sampling system, one can get the discrete value of the signal $f(x)$ at the sampling points $\{x_0, x_1, ..., x_{n-1}\}$. Here, the sampling points are assumed to be uniformly distributed on the interval of $[x_0, x_{n-1}]$, and the discrete value of the signal $f(x)$ at the sampling points are denoted as $f_0, f_1, ..., f_{n-1}$. One can calculate the frequency spectrum $F(\omega)$ as

$$F(\omega_k) = F_k = \sum_{j=0}^{n-1} f_j e^{-i2\pi kj/n} \tag{1}$$

where $k = 0, 1, \ldots, n-1$, and $\omega_k$ is evenly distributed. The adjacent interval between $\omega_k$ depends on the sampling interval. $F(\omega)$ is conjugate symmetric when the signal is real-valued. Here, $\{F_k\}$ is divided into $m$ segments of length $\Delta\omega$ in order, in which $n = m\Delta\omega$.

$$\{F_k\}_{k=0,1,\ldots,n-1} = \bigcup_{i=0}^{m-1} S_i \tag{2}$$

where

$$S_i = \{F_k\}_{k=i\Delta\omega, i\Delta\omega+1, \ldots, (i+1)\Delta\omega-1} \tag{3}$$

In the following, we call $S_i$ the $i$th segment of $F(\omega)$. Considering the conjugate symmetry, only half part of $\{F_k\}$ needs fitting. Further considering that the sampling frequency is much larger than the bandwidth of the signal ($\omega_b = b\Delta\omega$), Eq. (2) can be approximated by

$$\{F_k\}_{k=0,1,\ldots,n/2} \approx \left(\bigcup_{i=0}^{b} S_i\right) \cup \left(\bigcup_{i=b+1}^{m/2} O_i\right) \tag{4}$$

where

$$O_i = \underbrace{\{0, 0, \ldots, 0\}}_{\Delta\omega} \tag{5}$$

and

$$F_{n-k} = F_k^* \quad k = 1, 2, \ldots, (n-1)/2 \tag{6}$$

in which * stands for the conjugate operation. For each slice $S_i$, a neural network is used to approximate the information $T_i$ it contains. More precisely,

$$T_i(k) \approx \begin{cases} F_k & k = i\Delta\omega, i\Delta\omega+1, \ldots, i\Delta\omega+\Delta\omega-1 \\ 0 & \text{otherwise} \end{cases} \tag{7}$$

where $i = 1, 2, \ldots, b$. One can see from Eq. (7) that, different from PhaseDNN [11], in the proposed method, each neural network $T_i$ is used to "memorizes" the discrete data in the frequency domain.

After obtaining these trained neural networks and their predictions $\{T_i\}$, the following concatenation operation is used to obtain the approximation of $F(\omega)$.

$$F_k \approx T(k) = \begin{cases} \sum_{i=0}^{b} T_i(k) & k = 0, 1, ..., (n-1)/2 \\ \sum_{i=0}^{b} T_i^*(n-k) & k = (n+1)/2, (n+1)/2+1, ..., n-1 \end{cases} \quad (8)$$

Finally, each sampling signals $f_j$ can be approximated by using the following inverse FFT (IFFT),

$$f_j \approx \frac{1}{n} \sum_{k=0}^{n-1} T(k) e^{i2\pi jk/n} \quad (9)$$

Let us compare the recent PhaseDNN method and the proposed PFF-DNN method from ease of operation. In PhaseDNN, $2m$ convolutions are required to get all frequency selected signals. Then, $2m$ frequency shifts and $2m$ inverse frequency shifts are also involved. In comparison, the proposed method is more straightforward, and only one FFT and one IFFT are required. No frequency selection, frequency shifts, nor inverse shifts is involved.

In the PhaseDNN method, DNNs are used to fit continuous signals in the time domain. On the contrary, in the proposed method, DNNs are used to "memorizes" discrete data points in the frequency domain. For many commonly used signals (including the signals recommended in Ref. [11] to demonstrate the effectiveness of the PhaseDNN method), detailed numerical comparisons are conducted to show that the neural network "memorize" such discrete frequency values much faster.

## 3. Numerical Experiments

### 3.1. Experimental Setting

Six typical signal analyses are performed to evaluate the fitting performance of our PFF-DNN method, accompanied by a comparison with the existing PhaseDNN method. These signals to be analyzed include some periodic and non-periodic analog signals in the form of explicit functions and complex signals described by the classical Mackey-Glass differential dynamic system. For each signal, some detailed comparative comparisons were conducted to describe the effectiveness and efficiency of different methods for fitting the broadband signals with the high-frequency components. The detailed comparison includes 1) convergence curves (the changes of the loss $R$ with respect to the number of updates $N$ of the neural network weights), 2) the convergence process of different frequency bands, 3) the influence of different $\Delta\omega$ on the convergence speed, and 4) the influence of different updating times $N$ on the fitting accuracy.

Before showing the experimental results, we first introduce the neural network model and training-related parameters for all subsequent experiments. The neural network structure used here is consistent with the network used in [11] (i.e., the 1-40-40-40-40-1 fully connected neural network). Each layer (except the output layer) uses the 'ELU' activation function [13]. The neural network training adopts the Adam optimizer [14], and the learning rate is 0.0002. The length of each training batch is 100. For the method in [11], 5000 training data and 5000 testing data are evenly distributed across the domain. For the method proposed in this paper, each slice's length of the training sample depends on the slice's length in the frequency domain. For both methods, we record the approximation error in rooted mean squared error (RMSE) when $\Delta\omega$ of the slice in the frequency domain is 11, 21, 31, 41, or 51, and when the total number of updating is from 1 to 10000. The $\Delta\omega$ used in the literature [11] is 11. So we gradually increase $\Delta\omega$ from this value. When $\Delta\omega$ increases to infinite, both methods will degenerate to a vanilla fitting method. Therefore, we limit the upper limit of $\Delta\omega$ to 51.

## 3.2. Signals with Explicit Expression

### 3.2.1. Sine on Polynomial

Here, the signal of sine on polynomial is used to test the neural network's approximation accuracy, which is similar to the one used in the literature [15]. It is described by Eq. (10), whose shape and the corresponding frequency spectrum are shown in Fig. 4. The signal consists of a high-frequency periodic component and a low-frequency non-periodic component.

$$f(x) = 0.1x^3 - 0.1x^2 - 0.5x + 0.3 + \sin(50x) \tag{10}$$

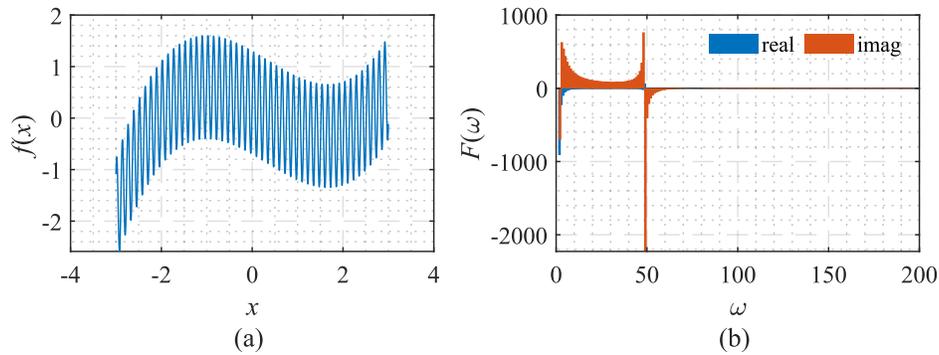

**Fig. 4.** Signal described by Eq. (10) with its frequency spectrum.

As shown in Fig. 5, only the low-frequency component can be approximated if we use the vanilla fitting method. No matter how large $N$ is, the high-frequency component cannot be fitted.

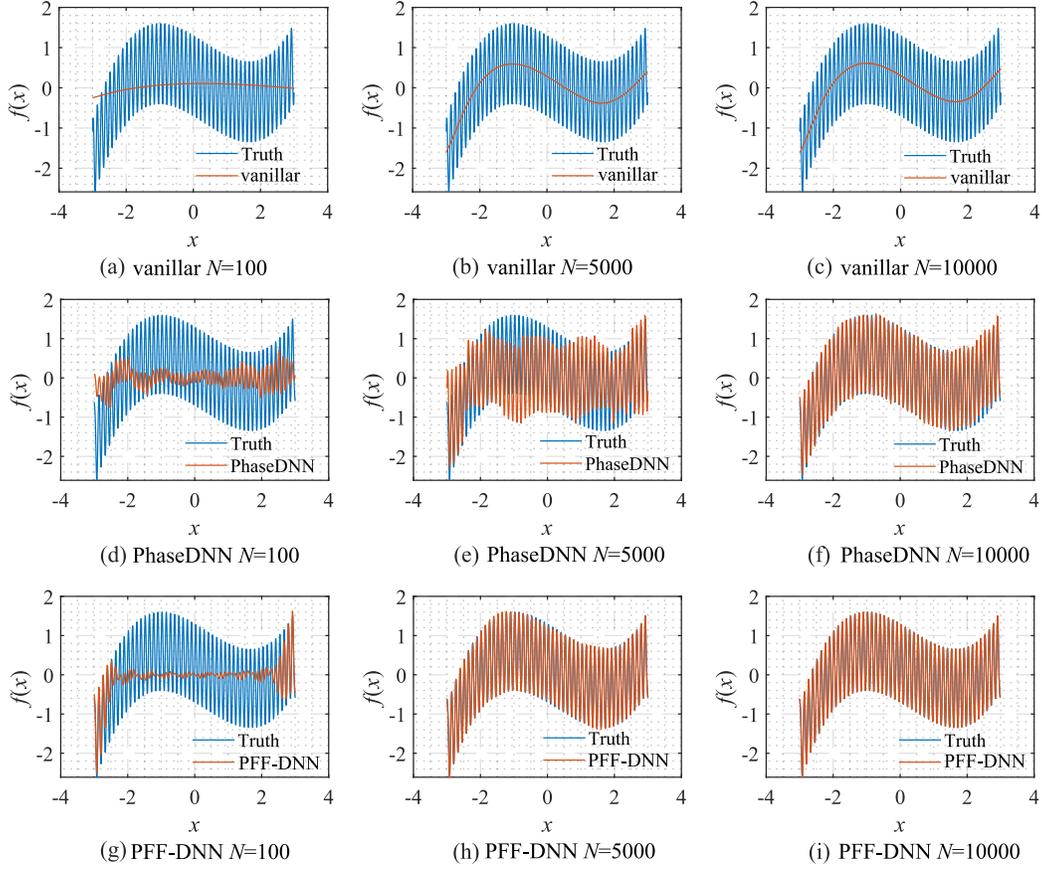

**Fig. 5.** Function fitting using different methods. Vanilla fitting result: (a) *N*=100, (b) *N*=5000, (c) *N*=10000; PhaseDNN fitting result: (d) *N*=100, (e) *N*=5000, (f) *N*=10000; PFF-DNN fitting results: (g) *N*=100, (h) *N*=5000, (i) *N*=10000.

On the contrary, if PhaseDNN or PFF-DNN is used, the signal can be fitted very well. As mentioned earlier, a larger $\Delta\omega$ indicate fewer neural networks used for fitting, which is desirable. For the current example, the bandwidth of the signal is 50 Hz. When $\Delta\omega = 11$, to approximate the signal precisely, one needs to use at least five networks for fitting. In comparison, when $\Delta\omega = 51$, one only needs two networks for fitting, which saves 80% of computation resources. However, an increase in $\Delta\omega$ will make the fitting harder.

Fig. 6 shows how the convergence process of PhaseDNN and PFF-DNN changes with $\Delta\omega$. We can see that both algorithms converge fast when $\Delta\omega$ is small. However, when $\Delta\omega$ increases, the convergence speed of both methods slows down. Note that, compared with PhaseDNN, the PFF-DNN algorithm is less sensitive to the increase of $\Delta\omega$. In detail, for PhaseDNN, when $\Delta\omega=11$, the network updates 10,000 times to reach convergence. However, when $\Delta\omega=41$, it is difficult to converge even if it is trained 100,000 times. In contrast, for PFF-DNN, when $\Delta\omega=41$, it converges within only 2000 updates. That is to say, when $\Delta\omega$ triples, the convergence time only doubles. This is desirable in practical applications.

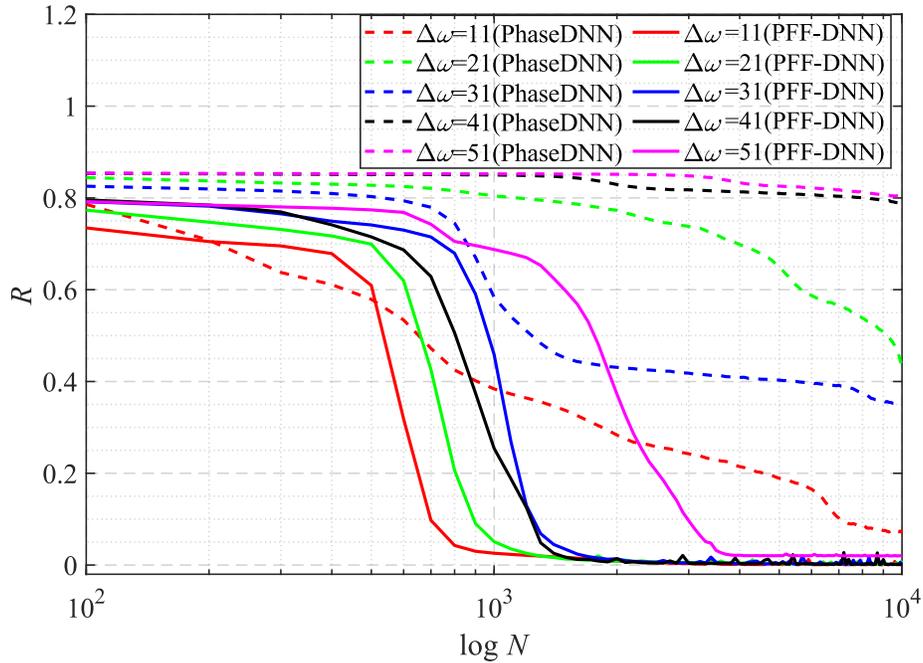

**Fig. 6.** The influence of $\Delta\omega$ on the convergence process.

Fig. 7 shows the fitting results of different frequency bands. Since the frequency spectrum for the neural network to approximate is not overcomplicated, the performance of PFF-DNN is better than that of Phase-DNN.

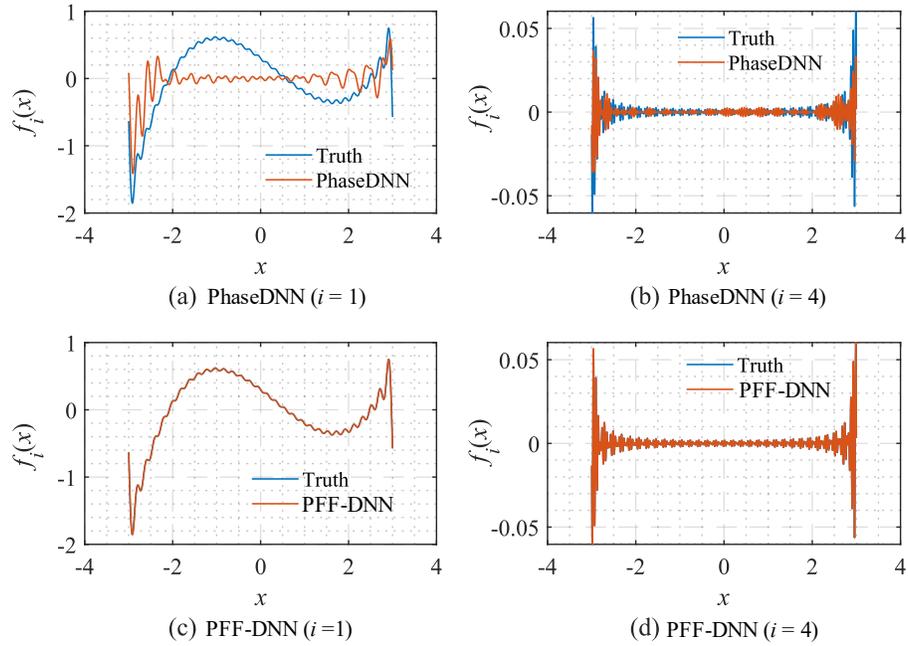

**Fig. 7.** Comparison between the fitting results from PhaseDNN and PFF-DNN for different frequency bands. The fitting results after neural networks are updated 5000 times given $\Delta\omega$=31: (a) First frequency band (PhaseDNN), (b) Fourth frequency band (PhaseDNN); (c) First frequency band (PFF-DNN); (d) Fourth frequency band (PFF-DNN ).

Fig. 8 shows the influence of $\Delta\omega$ on the algorithm performance. As shown in Fig. 8(c) and (f), when $\Delta\omega$=11, both methods can fit the signal well. However, when $\Delta\omega$ increases to 31, PhaseDNN begins to show fitting errors. When $\Delta\omega$=51, the PhaseDNN method can hardly fit the signal effectively. In comparison, the PFF-DNN method can still accurately fit the signal.

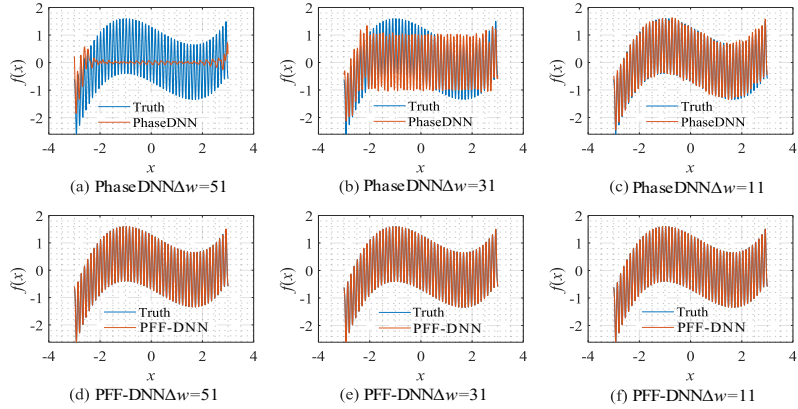

**Fig. 8.** The influence of $\Delta\omega$ on the algorithm performance. The fitting results of PhaseDNN after 10000 updates: (a) $\Delta\omega=51$, (b) $\Delta\omega=31$, (c) $\Delta\omega=11$; and fitting results of PFF-DNN after 10000 updates : (d) $\Delta\omega=51$, (e) $\Delta\omega=31$, (f) $\Delta\omega=11$.

Fig. 9 shows how the number of updating $N$ influences the algorithm performance. In the beginning, both algorithms do not fit the high-frequency components very well. The comparison between Fig. 9(b) and (e) shows that PFF-DNN converges faster compared with PhaseDNN. The red curves in Fig. 6 also confirm this point.

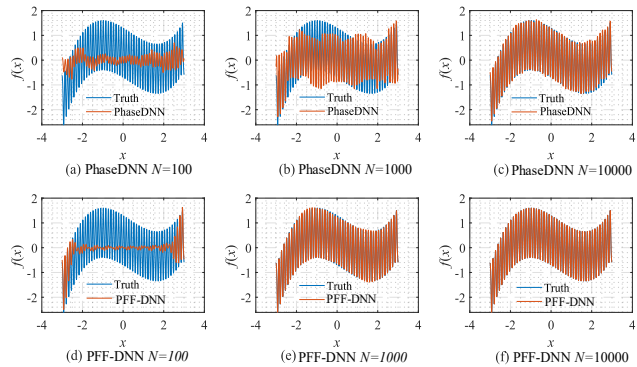

**Fig. 9.** The influence of $N$ on the algorithm performance. Fitting results of PhaseDNN when $\Delta\omega=11$: (a) $N=100$; (b) $N=1000$; (c) $N=10000$; Fitting results of PFF-DNN after 100 updates when $\Delta\omega=11$: (d) $N=100$; (e) $N=1000$; (f) $N=10000$.

*3.2.2. ENSO Signal*

The signal determined by the following equation is often used to approximate the ENSO data set [16]:

$$f(x) = 4.7\cos(2\pi x/12) + 1.1\sin(2\pi x/12)$$
$$+ 0.2\cos(2\pi x/1.7) + 2.7\sin(2\pi x/1.7)$$
$$+ 2.1\cos(2\pi x/0.7) + 2.1\sin(2\pi x/0.7)$$
$$- 0.5 \tag{11}$$

This signal is more complicated than the one described by Eq. (10). The shape of the signal and its corresponding frequency spectrum is shown in Fig. 10.

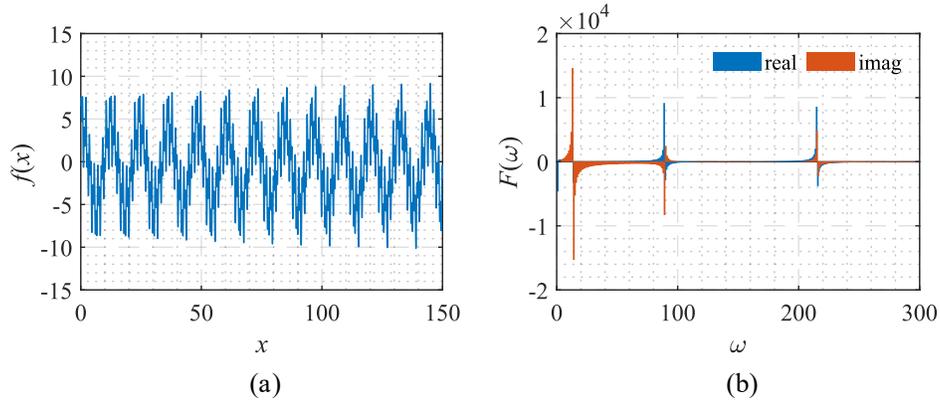

(a)     (b)

**Fig. 10.** Signal described by Eq. (11) with its frequency spectrum.

According to the trend of the solid red line and the red dashed line in Fig. 11, it can be seen that the convergence speed of PFF-DNN is about an order of magnitude larger. With the increase of $\Delta\omega$, the convergence time of PFF-DNN has only doubled. Thus, the larger $\Delta\omega$, the more training time PFF-DNN saves compared to PhaseDNN. Figs. 12 and 13 visually show the convergence process of the two networks.

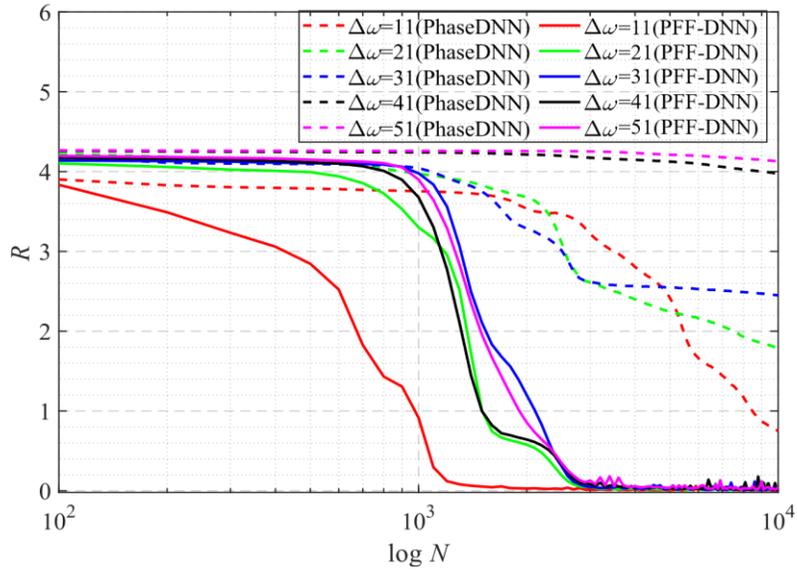

**Fig. 11.** The influence of $\Delta\omega$ on the convergence process.

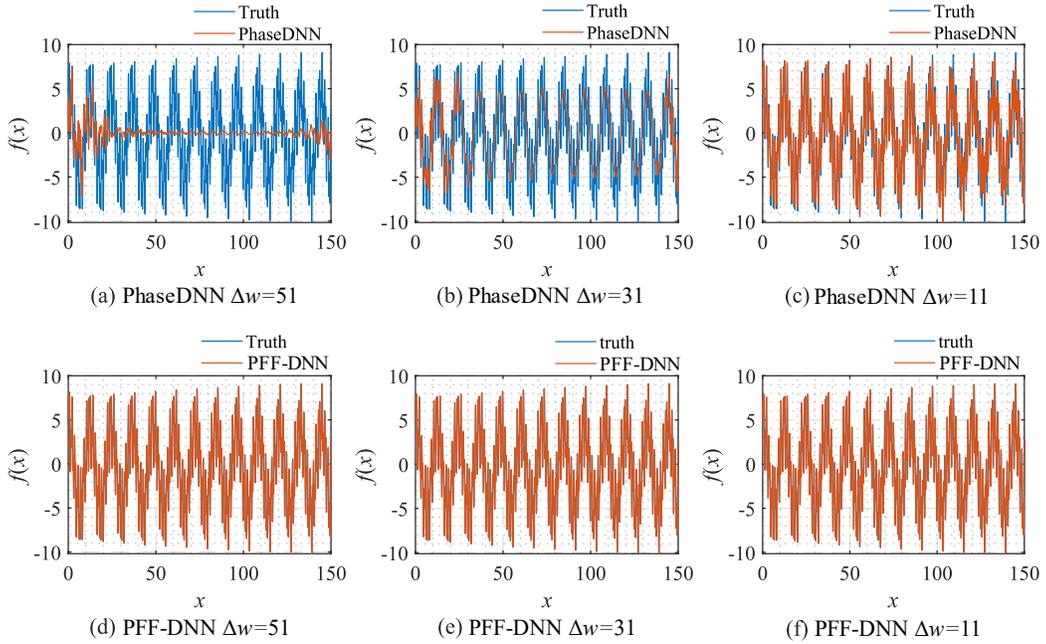

**Fig. 12.** The influence of $\Delta\omega$ on the algorithm performance. The fitting results of PhaseDNN after 10000 updates: (a) $\Delta\omega=51$, (b) $\Delta\omega=31$, (c) $\Delta\omega=11$; and fitting results of PFF-DNN after 10000 updates: (d) $\Delta\omega=51$, (e) $\Delta\omega=31$, (f) $\Delta\omega=11$.

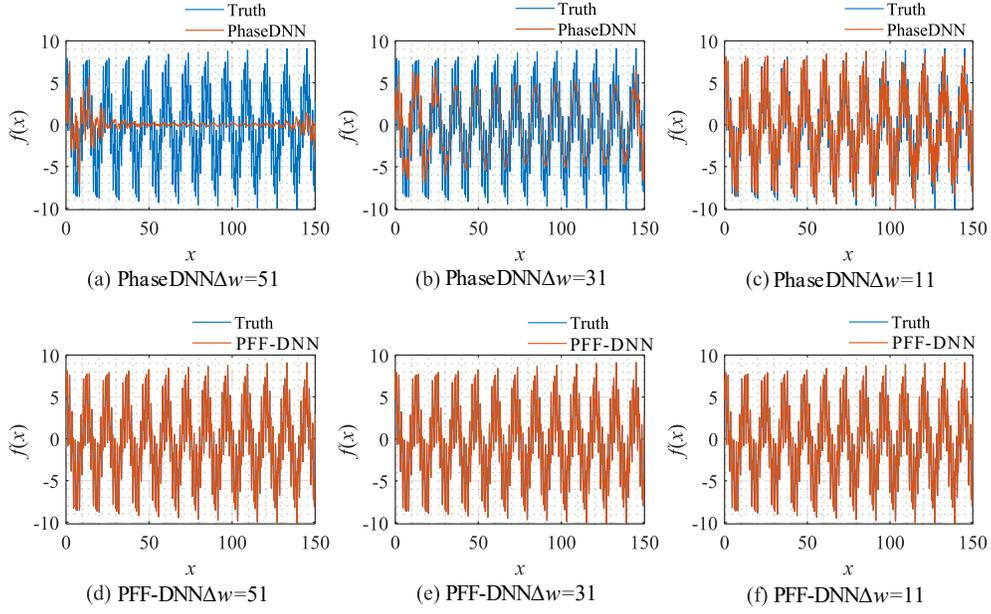

**Fig. 13.** The influence of *N* on the algorithm performance. Fitting results of PhaseDNN when $\Delta\omega$ =11: (a) *N*=100; (b) *N*=1000; (c) *N*=10000; Fitting results of PFF-DNN after 100 updates when $\Delta\omega$ =11: (d) *N*=100; (e) *N*=1000; (f) *N*=10000.

*3.2.3. Signal with Increasing Frequency*

Next, a signal with fast frequency changes is analyzed. This signal is often used in system identification [17], and its shape and frequency spectrum are shown in Fig. 14. The explicit expression of the signal used here is:

$$f(x) = \cos\left[\pi\left(f_0 + \frac{f_T - f_0}{T}x^3\right)x^3\right] \tag{12}$$

in which, *T*=1, $f_0$=0.01, and $f_T$ =50. We can see from Fig. 14(a) that as *x* increases, the oscillation frequency of the signal increases sharply. Fig. 14(b) shows that the amplitude and the oscillation frequency of the signal's spectrum gradually decrease.

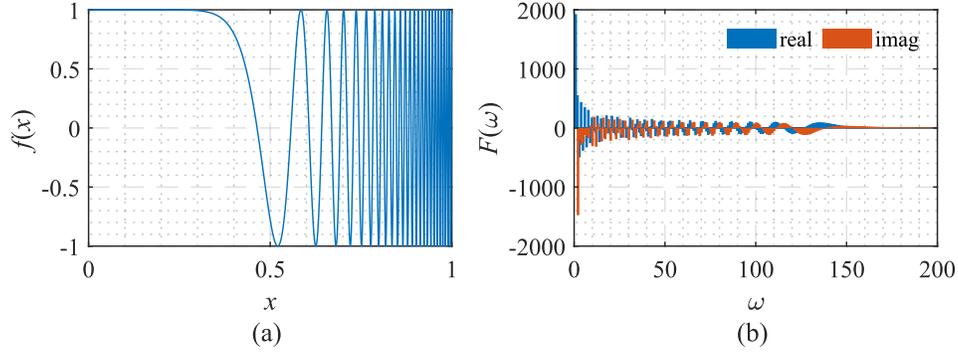

**Fig. 14.** Signal described by Eq. (12) with its frequency spectrum

From the curve in Fig. 15, we can see that when $\Delta\omega=11$, the efficiency and accuracy of PFF-DNN are better than that of PhaseDNN. When $\Delta\omega$ is increases, the convergence speed of both methods slows down rapidly. Note that the convergence speed of the PFF-DNN network is relatively faster. Figs. 16 and 17 visually compare the convergence process of the two methods.

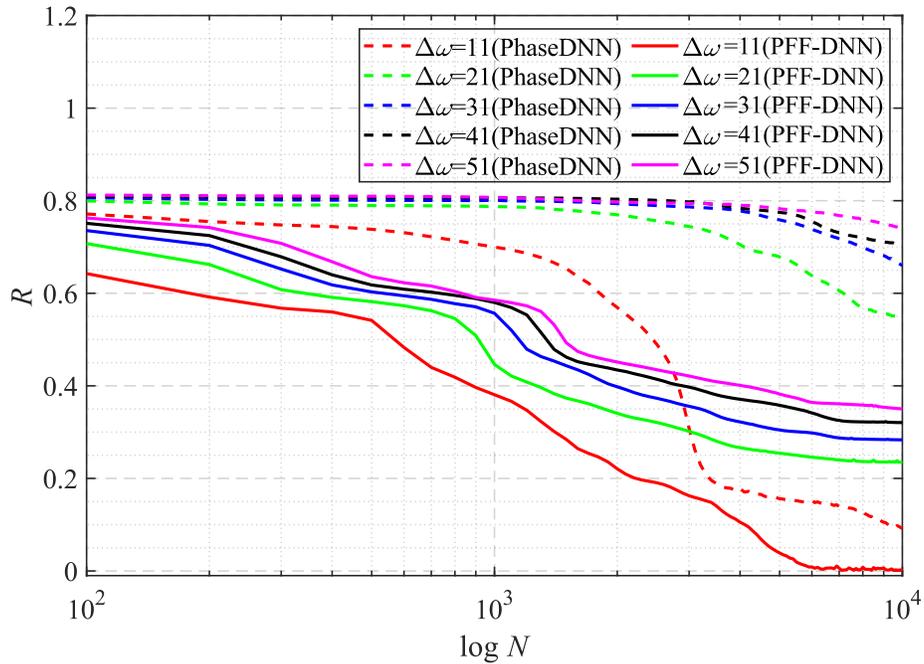

**Fig. 15.** The influence of $\Delta\omega$ on the convergence process.

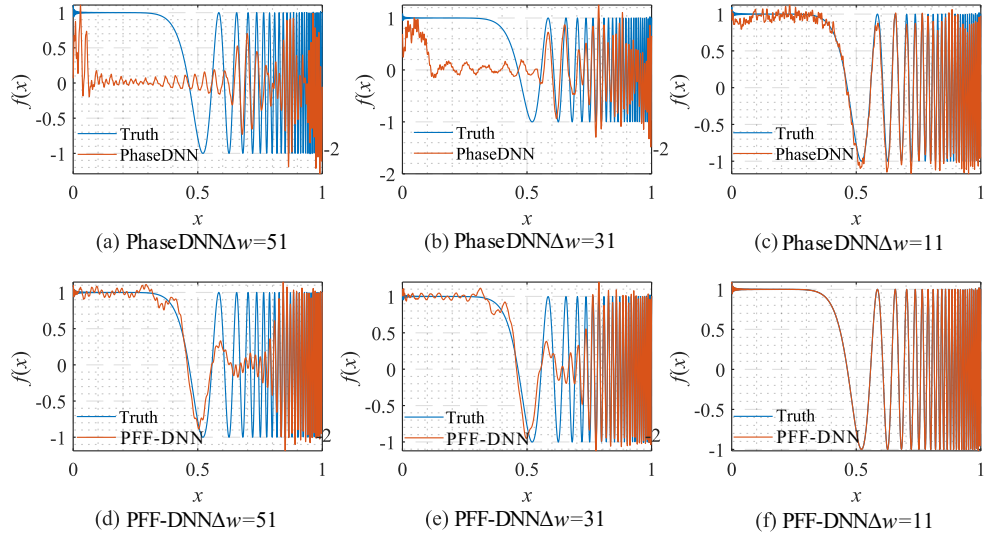

**Fig. 16.** The influence of $\Delta\omega$ on the algorithm performance. The fitting results of PhaseDNN after 10000 updates: (a) $\Delta\omega=51$, (b) $\Delta\omega=31$, (c) $\Delta\omega=11$; and fitting results of PFF-DNN after 10000 updates: (d) $\Delta\omega=51$, (e) $\Delta\omega=31$, (f) $\Delta\omega=11$.

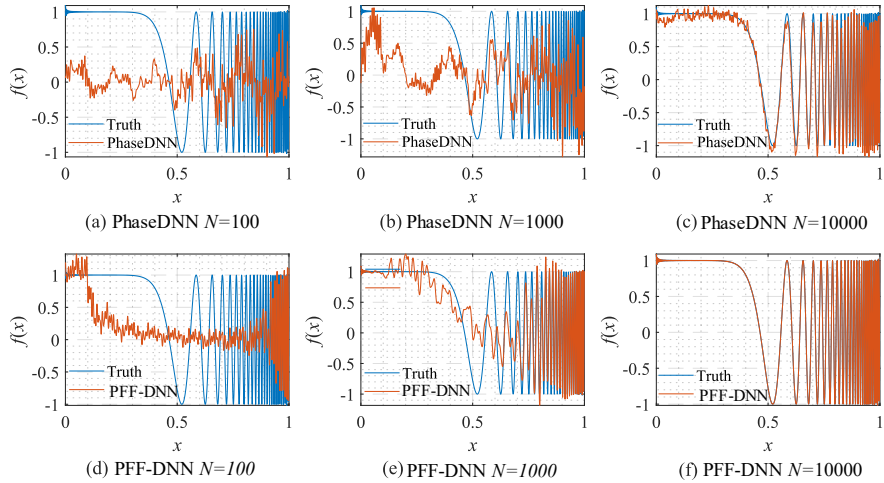

**Fig. 17.** The influence of $N$ on the algorithm performance. Fitting results of PhaseDNN when $\Delta\omega=11$: (a) $N=100$; (b) $N=1000$; (c) $N=10000$; Fitting results of PFF-DNN after 100 updates when $\Delta\omega=11$: (d) $N=100$; (e) $N=1000$; (f) $N=10000$.

## 3.2.4. Piecewise Signal

The piecewise signal is used in reference [11] to test neural networks' performance. The shape and its corresponding frequency spectrum are shown in Fig. 18. The signal is described by:

$$f(x) = \begin{cases} 10(\sin x + \sin 3x) & x \in [-\pi, 0] \\ 10(\sin 23x + \sin 137x + \sin 203x) & x \in [0, +\pi] \end{cases} \tag{13}$$

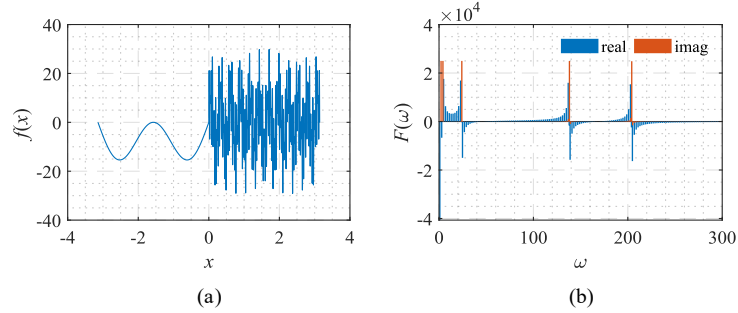

(a)          (b)

**Fig. 18.** Signal described by Eq. (13) with its frequency spectrum

The spectrum of this signal has several obvious spikes and many slight oscillations. It can be seen from Fig. 19 that for such signal, the accuracy of the proposed method is much less sensitive to changes in $\Delta\omega$. Figs. 20 and 21 visually compare the convergence process of the two networks.

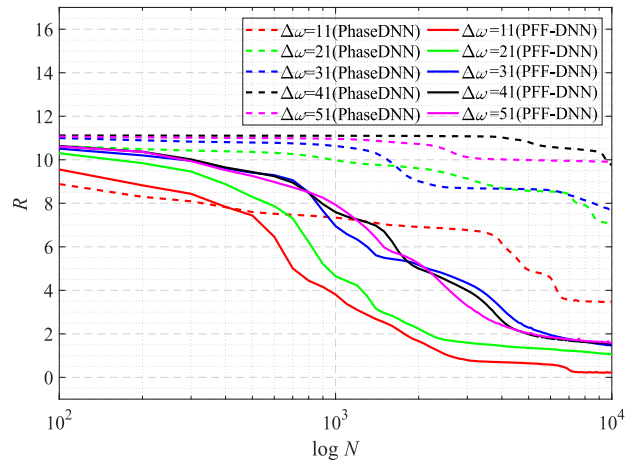

**Fig. 19.** The influence of $\Delta\omega$ on the convergence process.

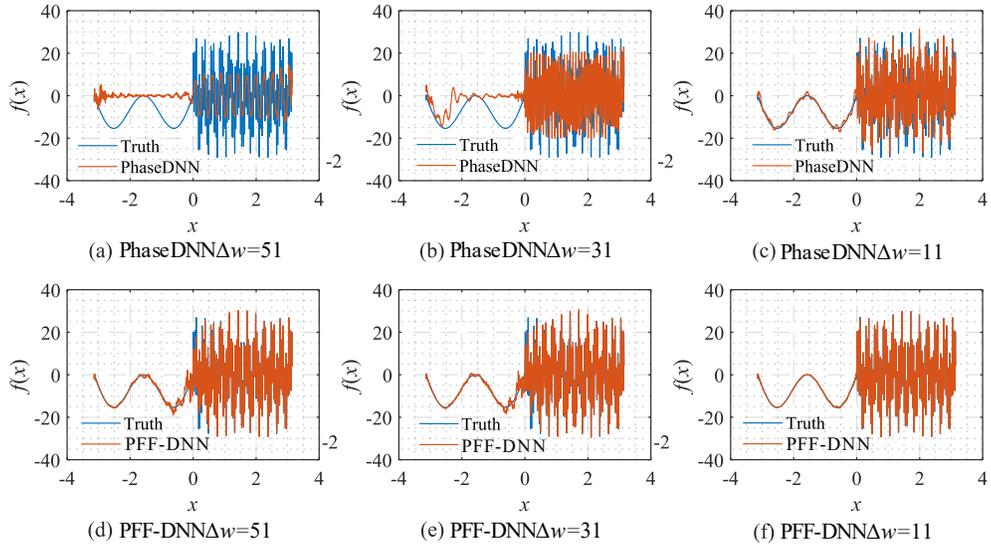

**Fig. 20.** The influence of $\Delta\omega$ on the algorithm performance. The fitting results of PhaseDNN after 10000 updates: (a) $\Delta\omega=51$, (b) $\Delta\omega=31$, (c) $\Delta\omega=11$; and fitting results of PFF-DNN after 10000 updates : (d) $\Delta\omega=51$, (e) $\Delta\omega=31$, (f) $\Delta\omega=11$.

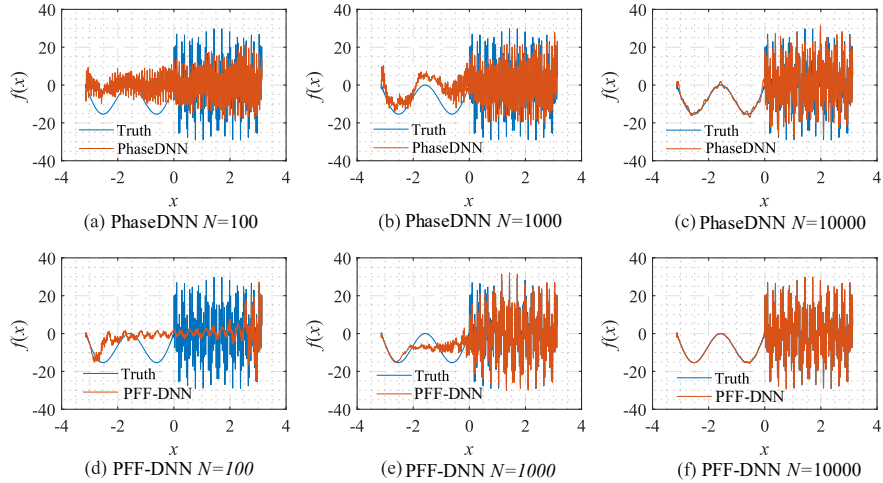

**Fig. 21.** The influence of $N$ on the algorithm performance. Fitting results of PhaseDNN when $\Delta\omega=11$: (a) $N=100$; (b) $N=1000$; (c) $N=10000$; Fitting results of PFF-DNN after 100 updates when $\Delta\omega=11$: (d) $N=100$; (e) $N=1000$; (f) $N=10000$.

## 3.2.5. Square Wave

The following example is used to test the proposed method's approximation accuracy on discontinuous signals such as square waves [11]. The shape and its corresponding frequency spectrum are shown in Fig. 22. The signal is described by:

$$f(x) = \sin(x) + \text{sign}\left[\sin(13x)\right] + \text{sign}\left[\sin(23x)\right] + \text{sign}\left[\sin(47x)\right] \tag{14}$$

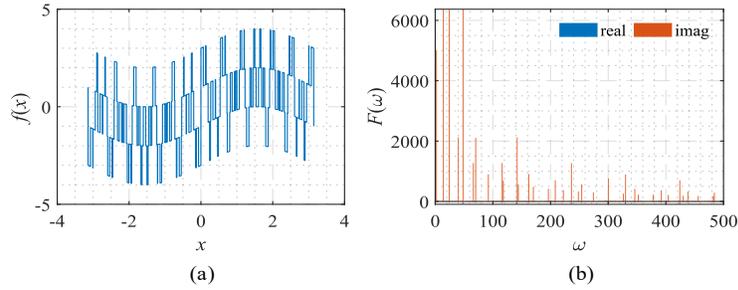

**Fig. 22.** Signal described by Eq. (13) with its frequency spectrum

The spectrum of this signal is composed of many irregular spikes. From Fig. 23, we can see that the increase in $\Delta\omega$ has a limited impact on PFF-DNN but has a significant impact on PhaseDNN. Figs. 24 and 25 visually compare the convergence process of the two networks.

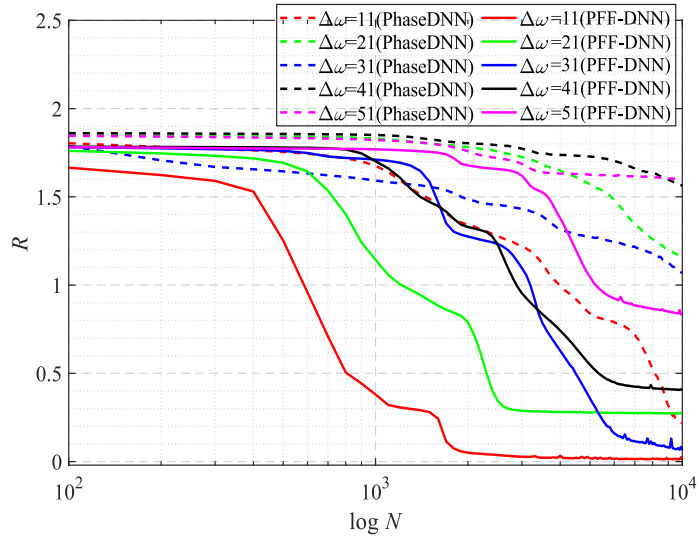

**Fig. 23.** The influence of $\Delta\omega$ on the convergence process.

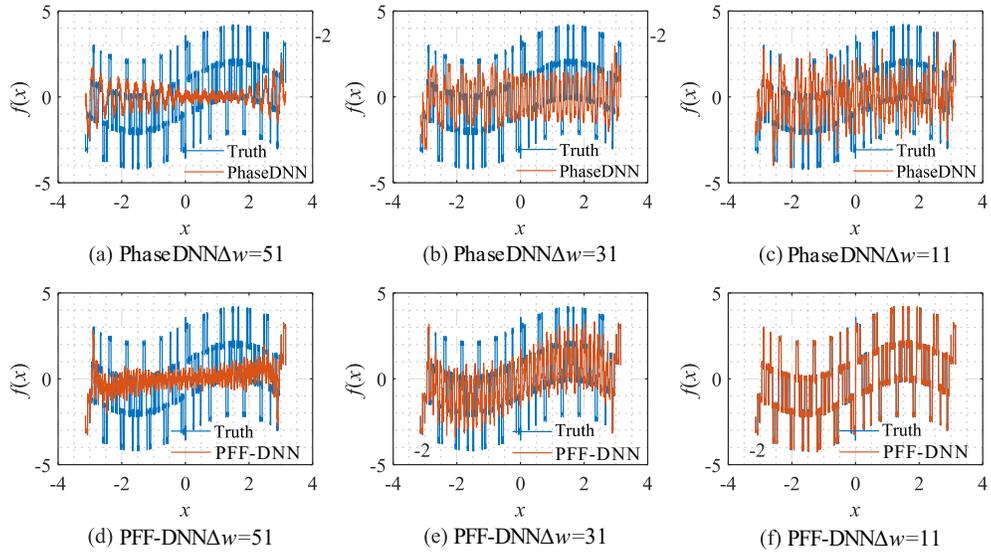

**Fig. 24.** The influence of $\Delta\omega$ on the algorithm performance. The fitting results of PhaseDNN after 10000 updates: (a) $\Delta\omega$ =51, (b) $\Delta\omega$ =31, (c) $\Delta\omega$ =11; and fitting results of PFF-DNN after 10000 updates : (d) $\Delta\omega$ =51, (e) $\Delta\omega$ =31, (f) $\Delta\omega$ =11.

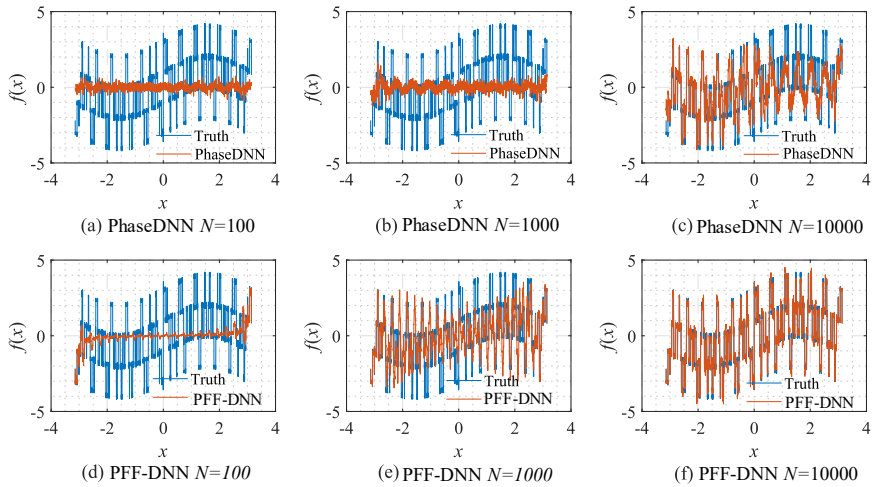

**Fig. 25.** The influence of $N$ on the algorithm performance. Fitting results of PhaseDNN when $\Delta\omega$ =11: (a) $N$=100; (b) $N$=1000; (c) $N$=10000; Fitting results of PFF-DNN after 100 updates when $\Delta\omega$ =11: (d) $N$=100; (e) $N$=1000; (f) $N$=10000.

## 3.3. Dynamic System

The Mackey-Glass equation is a time-delayed differential equation first proposed to model white blood cell production [18]. This equation has been used to test signal approximations in many related works [19-21]. It is described by the solution of the following lagged differential equation:

$$\dot{y}(x) = \frac{0.2 y(x-30)}{1 + y^{10}(x-30)} - 0.1 y(x) \tag{15}$$

The shape of the signal and its corresponding frequency spectrum are shown in Fig. 26.

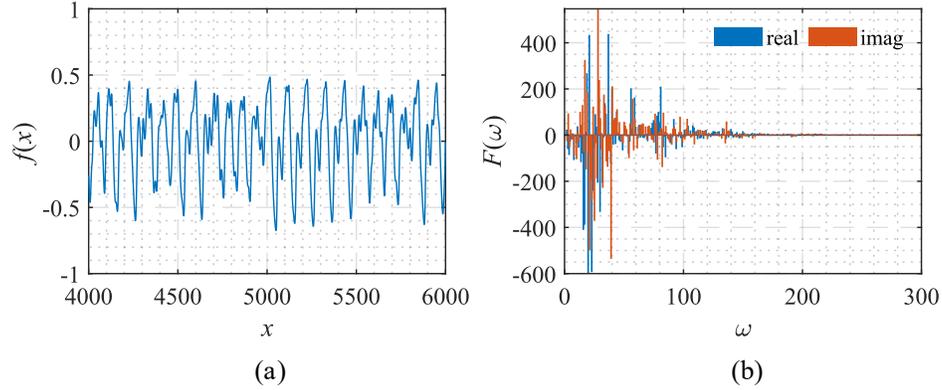

(a)          (b)

**Fig. 26.** Signal described by Eq. (13) with its frequency spectrum

As one can see, this signal's frequency spectrum is more complicated. It is distributed irregularly within the interval of [0, 200]. Thus, this problem is relatively hard for both PhaseDNN and PFF-DNN.

We can see from Fig. 27 that when $\Delta\omega = 21$ and $\Delta\omega = 41$, the training advantage of PFF-DNN over PhaseDNN is relatively obvious. However, when $\Delta\omega = 11$ and $\Delta\omega = 51$, the advantage is less obvious. Figs. 28 and 29 show that in the middle of training, the convergence speed of PFF-DNN is also slightly better.

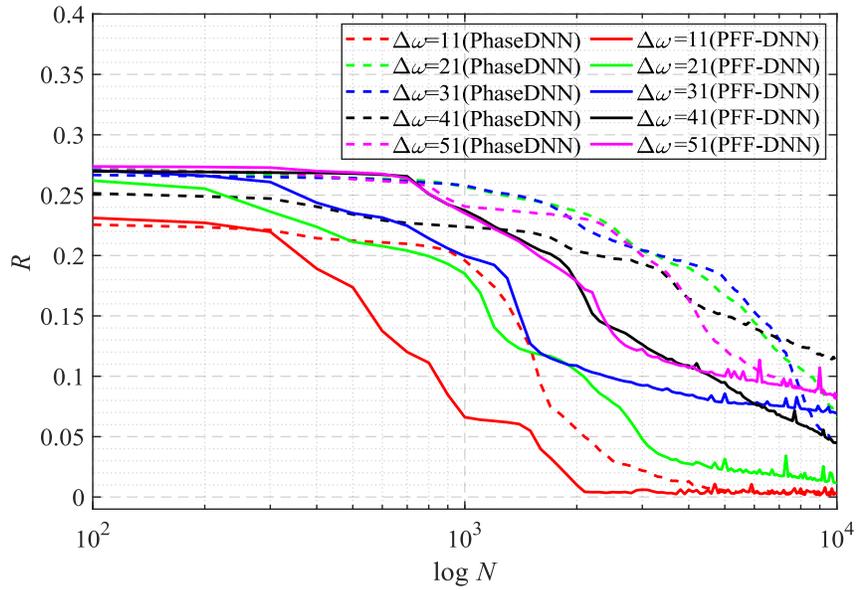

**Fig. 27.** The influence of $\Delta\omega$ on the convergence process.

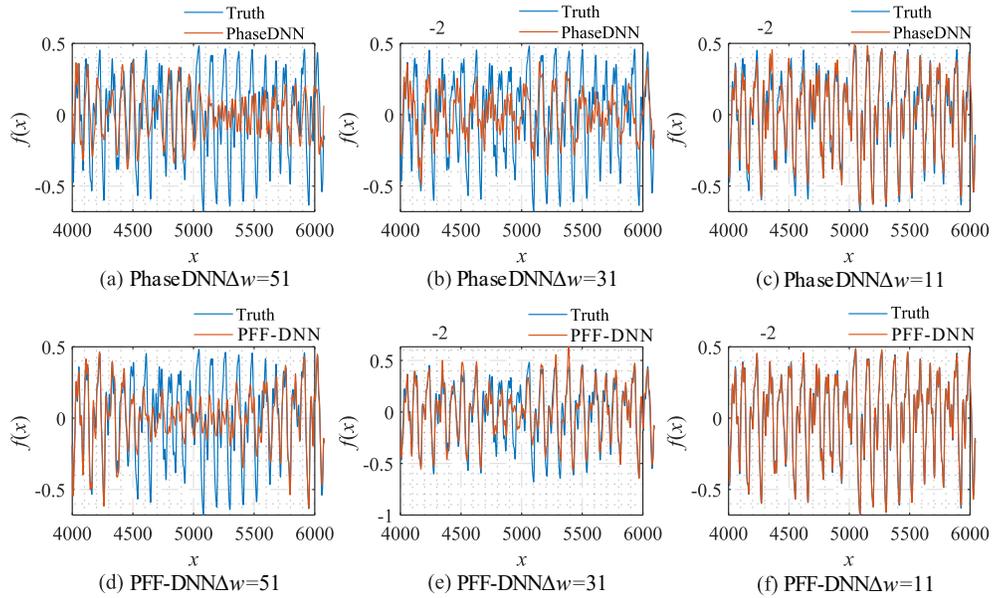

**Fig. 28.** The influence of $\Delta\omega$ on the algorithm performance. The fitting results of PhaseDNN after 10000 updates: (a) $\Delta\omega=51$, (b) $\Delta\omega=31$, (c) $\Delta\omega=11$; and fitting results of PFF-DNN after 10000 updates: (d) $\Delta\omega=51$, (e) $\Delta\omega=31$, (f) $\Delta\omega=11$.

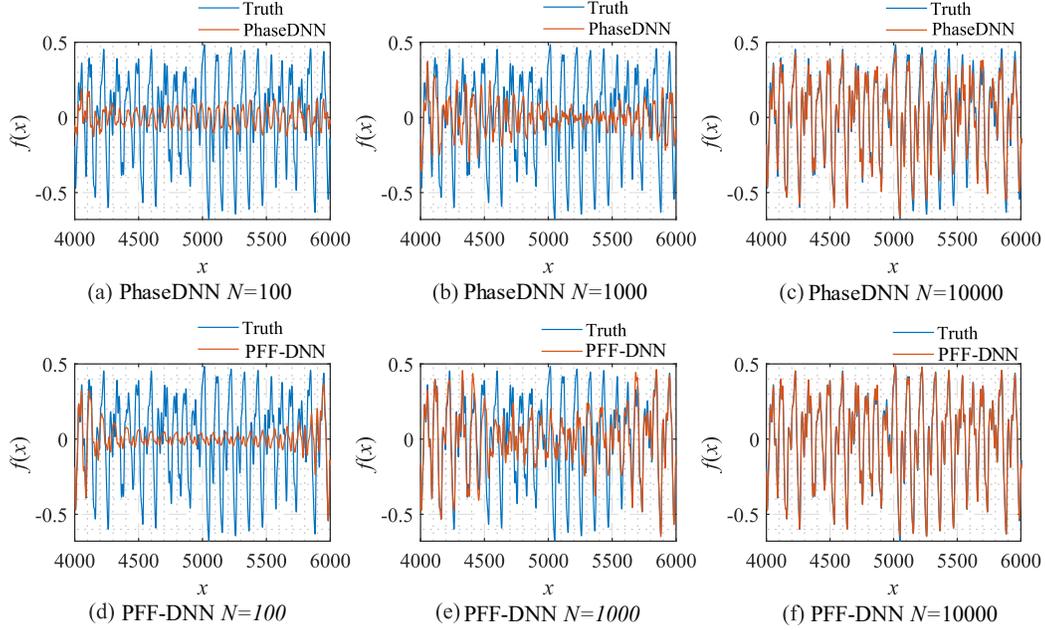

**Fig. 29.** The influence of *N* on the algorithm performance. Fitting results of PhaseDNN when $\Delta\omega$ =11: (a) *N*=100; (b) *N*=1000; (c) *N*=10000; Fitting results of PFF-DNN after 100 updates when $\Delta\omega$ =11: (d) *N*=100; (e) *N*=1000; (f) *N*=10000.

From this example, we can see that when the signal's frequency spectrum is relatively complicated, the performance gain of PFF-DNN is not significant. Therefore, we believe that PFF-DNN and PhaseDNN may have their advantages for different applications.

### *3.4. Further Discussions*

This section explains why PFF-DNN performs better in the above examples from the perspective of spectral bias of neural networks. To this end, Fig. 30 shows how PhaseDNN and PFF-DNN approximate the third frequency band when $\Delta\omega$=11 during the training process of the example in Section 4.2.3. It can be seen from Fig. 30(a) and (b) that PhaseDNN essentially handles a continuous signal in the time domain, while PFF-DNN essentially interpolates a frequency-domain discrete signal. Fig. 30(d) and (f) show the approximate results from PFF-DNN as a continuous function, which means that PFF-DNN interpolates the discrete spectrum with a smooth curve. Compared with Fig. 30(c) and (e), it can be seen that the curve approximated by PhaseDNN oscillates more frequently in the time domain. Thus, the neural network is more difficult to learn the signals in the time domain by considering the spectral bias. This discussion may explain why PFF-DNN performs better than PhaseDNN for these examples.

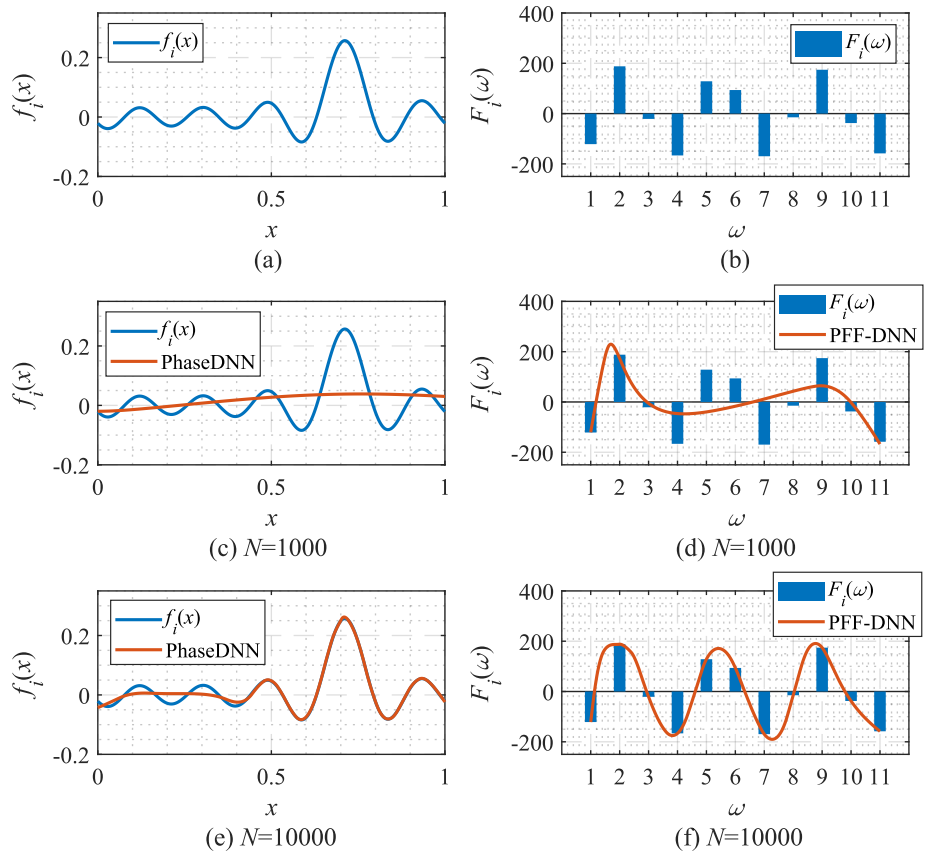

**Fig. 30.** Approximation performance during training. (a) The third frequency band of the fitted function (the real part after frequency-shift); (b) The spectrum (real part) of the function in (a); (c) Fitting results of PhaseDNN after 1000 updates; (d) Fitting results of PFF-DNN after 1000 updates; (e) Fitting results of PhaseDNN after 10000 updates; (f) Fitting results of PFF-DNN after 10000 updates.

## 4. Conclusions

Excessive training overhead is inevitable when the DNN or latest PhaseDNN method is used to fit a broadband signal. In order to overcome it, this paper proposes an alternative method based on parallel fitting in the frequency domain by utilizing fast Fourier analysis of broadband signals and the spectral bias nature of neural networks. The detailed comparisons were carried out based on numerical experiments for six typical broadband signals, showing that PFF-DNN has advantages in both accuracy and efficiency. With significant advantages in approximating high-frequency components of complex signals, the PFF-DNN method is expected to become an alternative solution for broadband signal fitting.

**Acknowledgment**

The authors gratefully acknowledge the financial support from the National Natural Science Foundation of China under Grant No. 11802225, No. 61805185, and No. 51805397.

**References**

[1] I. GOODFELLOW, Y. BENGIO, A. COURVILLE AND Y. BENGIO, *Deep Learning*, MIT Press, Cambridge, 2016.

[2] W. RAWAT AND Z. WANG, *Deep convolutional neural networks for image classification: a comprehensive review,* Neural Comput., 29 (2017), pp. 2352–2449.

[3] A. GRAVES, A.R. MOHAMED AND G. HINTON, *Speech recognition with deep recurrent neural networks*, in 2013 IEEE International Conference on Acoustics, Speech and Signal Processing, IEEE, Vancouver, BC, 2013, pp. 6645–6649.

[4] A. VASWANI, N. SHAZEER, N. PARMAR, J. USZKOREIT, L. JONES, A. N. GOMEZ, L. KAISER AND I. POLOSUKHIN, *Attention is all you need,* (2017), pp. arXiv preprint arXiv:1706.03762.

[5] Z. M. FADLULLAH, F. TANG, B. MAO, N. KATO, O. AKASHI, T. INOUE AND K. MIZUTANI, *State-of-the-art deep learning: evolving machine intelligence toward tomorrow's intelligent network traffic control systems,* IEEE Commun. Surv. Tutor., 19 (2017), pp. 2432–2455.

[6] P. KIDGER AND T. LYONS, *Universal approximation with deep narrow networks*, in Proceedings of Thirty Third Conference on Learning Theory, A. Jacob and A. Shivani, eds., PMLR, Graz, Austria, 2020, pp. 2306–2327.

[7] Z. LU, H. PU, F. WANG, Z. HU AND L. WANG, *The expressive power of neural networks: a view from the width,* (2017), pp. arXiv preprint arXiv:1709.02540.

[8] G. CYBENKO, *Approximation by superpositions of a sigmoidal function,* Math. Control Signals Sys., 2 (1989), pp. 303–314.

[9] N. RAHAMAN, A. BARATIN, D. ARPIT, F. DRAXLER, M. LIN, F. HAMPRECHT, Y. BENGIO AND A. COURVILLE, *On the spectral bias of neural networks*, in International Conference on Machine Learning, PMLR, Graz, Austria, 2019, pp. 5301–5310.

[10] Z. Q. J. XU, Y. ZHANG, T. LUO, Y. XIAO AND Z. MA, *Frequency principle: fourier analysis sheds light on deep neural networks,* (2019), arXiv preprint arXiv:1901.06523.

[11] W. CAI, X. LI AND L. LIU, *A phase shift deep neural network for high frequency approximation and wave problems,* SIAM J. Sci. Comput., 42 (2020), pp. A3285–A3312.

[12] H. J. NUSSBAUMER, *Fast Fourier Transform and Convolution Algorithms,* Springer, Berlin, 1981.

[13] CLEVERT, D.-A., T. UNTERTHINER, AND S. HOCHREITER, *Fast and accurate deep network learning by exponential linear units (elus),* (2017), pp. arXiv preprint arXiv: 1511.07289.

[14] KINGMA, D.P. AND J. BA, *Adam: A method for stochastic optimization,* (2014), pp. arXiv preprint arXiv: 1412.6980.

[15] S. YANG, T. TING, K. L. MAN AND S. U. GUAN, *Investigation of neural networks for function approximation,* Procedia Comput. Sci., 17 (2013), pp. 586–594.

[16] D. KAHANER, C. MOLER AND S. NASH, *Numerical Methods and Software,* Prentice-Hall, Inc, Englewood Cliffs N.J, 1989.

[17] X. G. XIA, *System identification using chirp signals and time-variant filters in the joint time-frequency domain,* IEEE Trans. Signal Process., 45 (1997), pp. 2072–2084.

[18] M. C. MACKEY AND L. GLASS, *Oscillation and chaos in physiological control systems,* Science, 197 (1977), pp. 287–289.

[19]  L. YINGWEI, N. SUNDARARAJAN AND P. SARATCHANDRAN, *A sequential learning scheme for function approximation using minimal radial basis function neural networks,* Neural Comput., 9 (1997), pp. 461–478.

[20]  R. D. JONES, Y. C. LEE, C. W. BARNES, G. W. FLAKE, K. LEE, P. S. LEWIS AND S. QIAN, *Function approximation and time series prediction with neural networks*, in 1990 IJCNN International Joint Conference on Neural Networks, IEEE, San Diego, CA, USA, 1990, pp. 649–665.

[21]  S. WU AND M. J. ER, *Dynamic fuzzy neural networks-a novel approach to function approximation,* IEEE Trans. Syst. Man Cybernet. B (cybernetics), 30 (2000), pp. 358–364.